\documentclass{emulateapj}
\usepackage{apjfonts}

\newcommand{\mH}{H$_2$}

\slugcomment{Accepted for publication in ApJL}

\shorttitle{Gamma-ray Emission from GeV-bright SNRs}
\shortauthors{Y. Uchiyama et al.}

\begin{document}

\title{Gamma-ray Emission from Crushed Clouds in Supernova Remnants}


\author{Yasunobu Uchiyama\altaffilmark{1}, Roger D. Blandford\altaffilmark{1}, 
Stefan Funk\altaffilmark{1}, Hiroyasu Tajima\altaffilmark{2}, Takaaki Tanaka\altaffilmark{3}}
\email{uchiyama@slac.stanford.edu}
\altaffiltext{1}{SLAC National Accelerator Laboratory, 2575 Sand Hill Road M/S 29, Menlo Park, CA 94025, USA.}
\altaffiltext{2}{Solar-Terrestrial Environment Laboratory, Nagoya University, 
Furo-cho, Chikusa-ku, Nagoya 464-8601, Japan}
\altaffiltext{3}{Kavli Institute for Particle Astrophysics and Cosmology, Stanford University, 382 Via Pueblo Mall, MC 4060, Stanford, CA 94305-4060, USA}

\begin{abstract}
It is shown that  the radio and gamma-ray emission 
observed from newly-found ``GeV-bright" supernova remnants (SNRs) 
can be explained by a model, in which a shocked cloud and 
shock-accelerated cosmic rays (CRs) frozen in it are 
simultaneously compressed by the supernova blastwave as a result of 
formation of a radiative cloud shock. 
Simple reacceleration of pre-existing CRs is generally 
sufficient to power the observed gamma-ray emission 
through the decays of $\pi^0$-mesons produced in 
hadronic interactions between high-energy protons (nuclei) and gas 
in the compressed-cloud  layer. 
This model provides a natural account of the observed 
synchrotron radiation in SNRs W51C, W44 and IC~443 with flat radio spectral index, 
which can be ascribed to a combination of secondary and reaccelerated electrons and positrons. 
\end{abstract}


\section{Introduction}

Luminous extended GeV $\gamma$-ray emission associated with 
middle-aged supernova remnants (SNRs) 
has recently been unveiled by the Large Area Telescope (LAT) 
onboard the \emph{Fermi} Gamma-ray Space Telescope.
Specifically, SNRs W51C, W44,  IC~443, and W28  are spatially resolved with 
the \emph{Fermi} LAT \citep{FermiW51C,FermiW44,FermiIC443,FermiW28}. 
The four SNRs are interacting with molecular clouds, as evidenced e.g., by
1720 MHz OH maser emission.  
The radio and $\gamma$-ray emission from these SNRs share similar characteristics. 
The synchrotron radio emission has a large flux of 160--310 Jy at 1 GHz 
with flat spectral index of $\alpha \simeq 0.26\mbox{--}0.40$. 
The GeV $\gamma$-ray spectrum commonly exhibits a spectral break at around 1--10 GeV, 
and the luminosity ranges $L_\gamma = (0.8\mbox{--}9)\times 10^{35}\, \rm erg\ s^{-1}$ 
in the 1--100 GeV band. 
Other cloud-interating SNRs associated with the LAT sources, 
such as CTB~37A, also emit $\gamma$-rays  
at a luminosity of  $L_\gamma \sim 10^{35}\, \rm erg\ s^{-1}$ 
\citep{CastroSlane}.

A prototypical example of the GeV-bright SNRs is SNR W44. 
The radio continuum map of W44 
exhibits filamentary and sheet-like structures  of synchrotron radiation 
well correlated with the shocked \mH\ emission
\citep{Reach05,W44radio}. 
According to \citet{Reach05}, the bulk of the synchrotron radiation 
can be ascribed to 
a fast molecular shock of a velocity  $v_{\rm s} \sim 100\ \rm km\ s^{-1}$ advancing through 
 a molecular cloud of a preshock density $n_0 \sim 200\ {\rm cm}^{-3}$. 
By passage of the blastwave of the supernova remnant, 
the shocked molecular cloud forms a thin sheet due to radiative cooling.
The radio filaments are thought to come from the compressed zone behind the shock front. 

The synchrotron radio emission arising from such a  ``crushed cloud" 
 was modeled by \citet{BC82}. 
It was shown that reacceleration of 
pre-existing CR electrons at a cloud shock and subsequent adiabatic compression 
results in enhanced synchrotron radiation, capable of explaining the radio 
intensity from evolved SNRs. 
\citet{Bykov00} discussed a similar scenario, in which direct electron 
acceleration from the thermal pool is also invoked. 

The $\pi^0$-decay $\gamma$-ray 
emission should be enhanced in the crushed clouds in the same manner 
as the synchrotron radiation. 
In this Letter, we demonstrate that  
the newly-found $\gamma$-ray emission from middle-aged SNRs can be 
readily understood within the crushed cloud scenario, in which 
the $\gamma$-ray emission comes from shocked clouds overrun by SNR blastwaves. 
It should be emphasized that our model is 
essentially different from the scenarios adopted in 
the recent papers \citep[e.g.,][]{Fujita10,Torres10}, where 
molecular clouds in the vicinity of SNRs are assumed to be illuminated by 
 runaway CRs \citep{AA96,Gabici09}. While the GeV and TeV $\gamma$-rays 
outside the southern boundary of SNR W28 \citep{HESSW28,AGILE_W28,FermiW28} 
may be explained by such runaway CRs, we argue here that 
the luminous GeV $\gamma$-ray emission in the directions of 
cloud-interacting SNRs emerges from the radiatively-compressed clouds.

\section{The Model}\label{sec:model}

\subsection{Cloud Shock Structure}\label{sec:shock}

Let us consider a strong shock driven into a molecular cloud 
by the  high pressure behind a supernova blastwave. 
Using the number density of hydrogen nucleus in the preshock cloud, $n_0$, 
the preshock magnetic field 
 is described by  a dimensionless parameter $b$: 
\begin{equation}
B_0 = b \sqrt{(n_0/{\rm cm}^{-3})} \ \mu{\rm G}. 
\end{equation}
Zeeman measurements of self-gravitating molecular clouds show that 
$b$ is roughly constant from one cloud to another  with  $b\sim 1$ \citep{Crutcher99}. 
We are concerned primarily with a fast ($v_{\rm s} \ga 50\ \rm km\ s^{-1}$) $J$-type 
shock \citep{J-shock}, in which 
ambipolar diffusion and radiative cooling are unimportant in a shock dissipation  layer. 
The steady-state postshock structure of a fast molecular shock is 
described in \citet{HM89}, which is referred to as HM89 hereafter. 

The initial temperature immediately behind the shock front is 
$T_{\rm s} \simeq 3.2 \times 10^5\, v_{\rm s7}^2/{x_t}\, {\rm K}$,
where $v_{\rm s7} \equiv v_{\rm s} / (100\ \rm km\ s^{-1})$, and 
$x_t$ is the number of particles per hydrogen nucleus (HM89). 
The density rises from  a preshock density $n_0$ by a factor of $r_{\rm sh}=4$, 
in the strong shock limit. 
For $v_{\rm s7} \ga 1.2$, ionizing radiation produced in the immediate postshock 
layer is strong enough to fully predissociate and preionize the upstream cloud. 

As the gas radiatively cools downstream, the temperature decreases and 
the density increases. The compression of the cooling gas is limited by 
the magnetic pressure in the cases that we shall be considering. 
The density of the cooled gas, $n_m$, is determined as 
\begin{equation}
{n_m} \simeq 94\,  n_0 v_{\rm s7} b^{-1}, 
\end{equation}
which is obtained by equating $B_m^2 /8\pi$ with the
shock ram pressure $n_0 \mu_{\rm H} v_s^2$, where 
$B_m = \sqrt{2/3} (n_m/n_0) B_0$ is the compressed magnetic field, and 
$\mu_{\rm H}$ is 
the mass per hydrogen nucleus (HM89).
(We have assumed that the preshock magnetic field is randomly directed.)
The [\ion{O}{1}] line (63$\mu$m)
surface brightness of a face-on $J$-shock 
is  proportional to the particle flux into the shock, $n_0 v_s$. 
The peak surface brightness of the [\ion{O}{1}] line 
is  $1\times 10^{-3}\ \rm erg\ cm^{-2}\, s^{-1}\, sr^{-1}$ in SNR W44 \citep{RR96}, 
and a factor of 2 smaller in IC~443 \citep{Rho01}, indicating the particle flux 
of order 
$n_0 v_s  \sim 10^9\  \rm cm^{-2}\, s^{-1}$ (HM89) and therefore 
 $n_m \sim 9\times 10^3\, b^{-1}\ \rm cm^{-3}$. 

A certain column density, $N_{\rm cool}$, has to be transmitted by a shock 
for gas to cool down to $10^4$ K and become radiative. 
This occurs in a column density 
$N_{\rm cool} \simeq 3\times 10^{17} \ v_{\rm s7}^{4}\ \rm cm^{-2}$ 
for $v_{s7} = 0.6\mbox{--}1.5$  \citep{McKee87}. 
Recombination and photoionization by the ultraviolet radiation produced upstream 
are balanced in  $N_{\rm cool} < N < N_{\rm ion} \sim 10^{19}\ \rm cm^{-2}$.
Beyond $N_{\rm ion}$,  the ionizing photons are absorbed and 
molecular chemistry commences. 
By setting 
a typical elapsed time since shocked to be $t_c = t/2$,  
where $t = 10^4\, t_4\, {\rm yr}$ is the age of the remnant, and using  
 $n_{0,2} \equiv n_0/(100 \ \rm cm^{-2})$, 
the  column density of  the compressed cloud is written as 
$N_c = n_0 v_s t_c \simeq 1.5 \times 10^{20} n_{0,2} v_{s7} t_4 \ {\rm cm}^{-2}$. 

The bulk of synchrotron radio waves and $\gamma$-rays 
should be emitted in the 
compressed gas with a constant density $n_m$ (within a factor of 2) 
and magnetic field $B_m$. 
Note that  we idealize the shock as one-dimentional and ignore 
any effects, such as lateral compression, caused by the secondary shocks. 
The secondary shocks could affect the leakage of 
high-energy particles from the compressed cloud,  playing 
an indirect role in the $\gamma$-ray production. 
Also, the ultraviolet radiation 
produced in the secondary shocks may change the ionization level of the precursor 
of the main cloud shock.

\subsection{Shock Acceleration and Adiabatic Compression}

Following \citet{BC82},  we  consider a conservative case 
in which 
only pre-existing CRs are accelerated by the process of diffusive shock acceleration 
at a cloud shock. 
Suprathermal particles may be injected to the acceleration process 
at the shock front, despite slow acceleration and fast Coulomb losses. 
However, we shall show below that the shock acceleration of  pre-existing CRs alone 
appears to suffice as the origin of the observed $\gamma$-ray emission 
from the cloud-interacting SNRs, 
and therefore consider the reacceleration case only. 

Let $n_{\rm acc}(p)dp$ be the CR number density in $p\sim p+dp$, 
transmitted by a shock, and $n_{\rm GCR}(p)$ be the pre-existing ambient CR 
density. 
According to the theory of diffusive shock acceleration \citep{BE87}, 
for $p < p_{\rm br/max}$ (see below), 
\begin{equation}
n_{\rm acc}(p) = (\alpha +2 )\ p^{-\alpha}
 \int_{0}^{p} dp^{\prime} n_{\rm GCR}(p^{\prime})\ p^{\prime \, (\alpha-1)}, 
\end{equation}
where $\alpha = (r_{\rm sh}+2)/(r_{\rm sh}-1)$ and 
$r_{\rm sh}$ is the shock compression ratio, which is assumed to be  $r_{\rm sh}=4$. 

Assuming that the density of the Galactic CRs in the molecular cloud 
is same as that in the general interstellar medium, 
we adopt the Galactic CR proton spectrum of the form:
\begin{equation}
n_{\rm GCR,p}(p) = 4\pi J_p \beta ^{1.5} p_{0}^{-2.76}, 
\end{equation}
where $p_0 = p/({\rm GeV}/c)$, 
$J_p = 1.9\, \rm cm^{-2}\, s^{-1}\, sr^{-1}\, GeV^{-1}$, $\beta$ is the proton velocity 
in units of $c$. 
A low-energy cutoff at a kinetic energy of 50 MeV is applied.  
This spectrum lies in between 
\citet{BESS07} and \citet{Strong04} at 100 MeV. 
For the CR electron+positron spectrum, we use: 
\begin{equation}
n_{\rm GCR,e}(p) = 4\pi J_e p_{0}^{-2} 
\left( 1 + p_{0}^{2} \right)^{-0.55}, 
\end{equation}
extending down to a low-energy cutoff of 20 MeV at which point 
ionization losses in the Galaxy should make the spectrum flat. 
The normalization factor is 
 $J_e = 2\times 10^{-2}\, \rm cm^{-2}\, s^{-1}\, sr^{-1}\, GeV^{-1}$.

The preshock upstream gas is only partially ionized for $v_{s7} \la 1.2$. 
Recently, \citet{Malkov10} have proposed that strong ion-neutral collisions 
accompanying Alfv\'en wave evanescence lead to steepening of 
the spectrum of accelerated particles; 
 the slope of the particle momentum distribution 
becomes steeper by one power above $p_{\rm br} = 2eB_0 V_{\rm A}/c\nu_{\rm i-n}$, 
where $V_A$ is the Alfv\'en velocity and 
$\nu_{\rm i-n} \simeq 9\times 10^{-9}\, n_{\rm n,0} T_{4}^{0.4}\ {\rm s}^{-1}$ is 
the ion-neutral collision frequency. Here $n_{\rm n,0}$ denotes the density of 
neutrals in units of $\rm cm^{-3}$ and $T_{4}$ is the precursor temperature in 
units of $10^4$ K. 
Interestingly, the parameters of  \citet{Reach05} estimated for 
the radio filaments of W44, namely $v_{\rm s7} \sim 1$ and $n_{0,2} \sim 2$, 
together with $B_0 = 30\, \mu$G and $T_4 = 1$ 
predict $p_{\rm br} \sim 10\ {\rm GeV}/c$ using the precursor ionization fraction 
calculated by HM89.  This agrees with the break value measured  
by the \emph{Fermi} LAT \citep{FermiW44}. 
We  introduce spectral steepening to $n_{\rm acc}(p)$, 
by multiplying a factor of $p_{\rm br}/p$ above $p_{\rm br}$. 

Even if the preshock gas is fully ionized, spectral steepening due to 
a finite acceleration time would be unavoidable already in the \emph{Fermi} bandpass. 
The timescale of diffusive shock acceleration can be written as 
$t_{\rm acc} \simeq (10/3) \eta c r_g v_s^{-2}$, where $r_g = cp/eB_0$ is 
the gyroradius and $\eta \geq 1$ is the gyrofactor; $\eta \sim 1$ 
has been obtained in young SNRs like RX~J1713.7$-$3946 \citep{Uchi07}. 
Equating $t_{\rm acc}$ with $t_c$, the maximum attainable energy is obtained as 
$cp_{\rm max} \simeq 50\,  (\eta /10)^{-1} v_{s7}^2 B_{-5} t_4\ \rm GeV$, 
where $B_{-5} = B_0/(10^{-5}\ \rm G)$. 
A factor of $\exp [ -(p/p_{\rm max})]$ is multiplied 
with $n_{\rm acc}(p)$ to introduce the maximum attainable energy.

The high-energy particles accelerated at the shock 
experience further heating due to adiabatic compression, as 
the gas density  increases until the pressure is magnetically supported. 
 Each particle gains 
energy as $p \rightarrow s^{1/3}p$, where $s \equiv (n_m/n_0)/r_{\rm sh}$, 
and the density increases by a factor of $s$ \citep[see][]{BC82}. 
Therefore, the number density of accelerated and compressed CRs 
at the point where the density becomes $\sim n_m$ is
\begin{equation}
n_{\rm ad}(p) = s^{2/3} n_{\rm acc}(s^{-1/3}p) .
\end{equation}

\subsection{Evolution in the Compressed Region}
In the compressed region of the cloud with a constant density $\sim n_m$, 
high-energy particles suffer from 
energy losses such as Coulomb/ionization losses. 
Also, the production of secondaries in inelastic proton-proton collisions 
can be  a significant 
source of high-energy electrons and positrons, since 
the energy loss timescale due to the {\it pp} collisions,  
$t_{pp} \simeq 6\times 10^7\, (n_m/{\rm cm}^{-3})^{-1}\, \rm yr$, 
generally becomes comparable to $t_c$. 

Let $N_p(p,t)dp$ represent the number of protons in $p\sim p+dp$ 
integrated over the emission volume at time $t$. 
We employ a usual kinetic equation to obtain the proton spectrum $N_p(p,t)$: 
\begin{equation}
\frac{\partial N_p(p,t)}{\partial t} = \frac{\partial}{\partial p} 
\left[ b(p) N_p(p,t) \right]  + Q_p(p), 
\label{eq:kinetic}
\end{equation}
where $Q_p$ is the proton injection rate, and $b(p)= -\dot{p}$ 
represents the proton energy losses. 
Charged particles with  $p < p_{\rm max}$ are assumed to be effectively trapped 
within the compressed gas. 
The particle spectrum  for accelerated electrons, 
$N_e(p,t)$, and for secondary $e^\pm$ resulting  from hadronic interactions,
$N_{\rm sec}(p,t)$, are computed similarly.

The injection rates of primaries are related to $n_{\rm ad}(p)$: 
\begin{equation}
Q(p) = \frac{n_0fV}{n_{m}t_c} n_{\rm ad}(p),
\end{equation}
where $V=4\pi R^3/3$ is the SNR volume with a radius $R$, 
and $f$ is the preshock filling factor of the compressed cloud. 
The volume of the preshock cloud is then $fV$. 
For simplicity, 
we assume that $Q(p)$ for primaries is time-independent; 
$v_s$ is assumed to be constant and $p_{\rm max}$ is evaluated 
simply at $t = t_c$. 
The injection rate of secondary $e^\pm$, $Q_{\rm sec} (p_e,t)$,  is  determined by 
$N_p(p_p,t)$ following the prescription given by \citet{Kamae}. 

\begin{deluxetable}{lr}
\tabletypesize{\small}
\tablecaption{Model Parameters for SNR W44 \label{tbl:model}}
\tablewidth{0pt}
\tablehead{
\colhead{Parameters} & \colhead{Values}}
\startdata
Assumed SNR Dynamics &     \\
\cline{1-1}\\
Distance: $D$ & 2.9 kpc\tablenotemark{a} \\
Radius: $R$ &  12.5 pc\tablenotemark{a} \\
 & ($\theta = 15\arcmin$)  \\
Age: $t$  & 10000 yr\tablenotemark{a} \\
Explosion energy: $E_{51}$ &  5\tablenotemark{a} \\
\cline{1-2} \\
Preshock Cloud Parameters  &\\
\cline{1-1}\\
Density: $n_{0}$ & $200\ \rm cm^{-3}$\tablenotemark{a} \\
Filling factor: $f$ & 0.18 \\
Magnetic field: $B_0$ & $25\, \mu{\rm G}$ \\
\cline{1-2} \\
Dependent Parameters & \\
\cline{1-1}\\
Cloud shock velocity: $v_s$ & $100\ \rm km\, s^{-1}$\tablenotemark{a}  \\
Break momentum ($T_4 =2$): $p_{\rm br}$ & $7\ {\rm GeV}/c$ \\
Maximum momentum ($\eta = 10$): $p_{\rm max}$ & $122\ {\rm GeV}/c$ 
\enddata
\tablenotetext{a}{Taken from \citet{Reach05}.}
\end{deluxetable}

\subsection{SNR Dynamics}

It may be desirable to relate the cloud shock velocity with the physical parameters 
that describe the blastwave of the remnant. 
We consider a remnant in the Sedov stage\footnote{This implies $t < t_{\rm tr}$, 
where $t_{\rm tr}$ is  an age for transition to the radiative phase of SNR 
evolution \citep{Blondin98}:
$t_{\rm tr} \sim 3 \times 10^4 \ E_{51}^{4/17}\ n_{\rm a,0}^{-9/17} \ {\rm yr}$.}:
\begin{equation}
\label{eq:ST1}
R_{12.5} = \left( \frac{E_{51}}{n_{\rm a,0}} \right) ^{1/5} t_4 ^{2/5}, 
\end{equation}
where $R_{12.5} = R/(12.5\ \rm pc)$, $E_{51}$ denotes the 
kinetic energy released by the supernova in units of $10^{51}\ \rm erg$, 
and 
$n_{\rm a,0}= n_{\rm a}/(\rm cm^{-3})$ represents the ambient (intercloud) density.
The blastwave velocity is $v_{\rm b} = 0.4R/t$. 
When a molecular cloud is struck by the blastwave, a strong shock is driven into the cloud 
with a shock velocity of 
\begin{eqnarray}
v_{\rm s} & = & k \left( \frac{n_{\rm a}}{n_0} \right) ^{1/2} v_{\rm b}, \nonumber \\
               & \simeq & 65\  n_{0,2}^{-1/2} \, E_{51}^{1/2} \, R_{12.5}^{-3/2}\ {\rm km\ s}^{-1},
\label{eq:vs}
\end{eqnarray}
where  $k \simeq 1.3$ is adopted. 
Note that $k$ depends weakly on $v_{\rm s}/v_{\rm b}$ \citep{MC75}, 
ranging $k=1\mbox{--}1.5$ in the circumstances of our interest. 
To drive a fast molecular shock of $v_{\rm s7} \ga 0.5$ into a molecular cloud, 
a condition of 
\begin{equation}
\label{eq:n0}
n_{0,2}  \la k^2\, E_{51}\, R_{12.5}^{-3}
\label{eq:condition}
\end{equation}
must be satisfied. 

\begin{figure}[htbp] 
 \epsscale{1.15}
\plotone{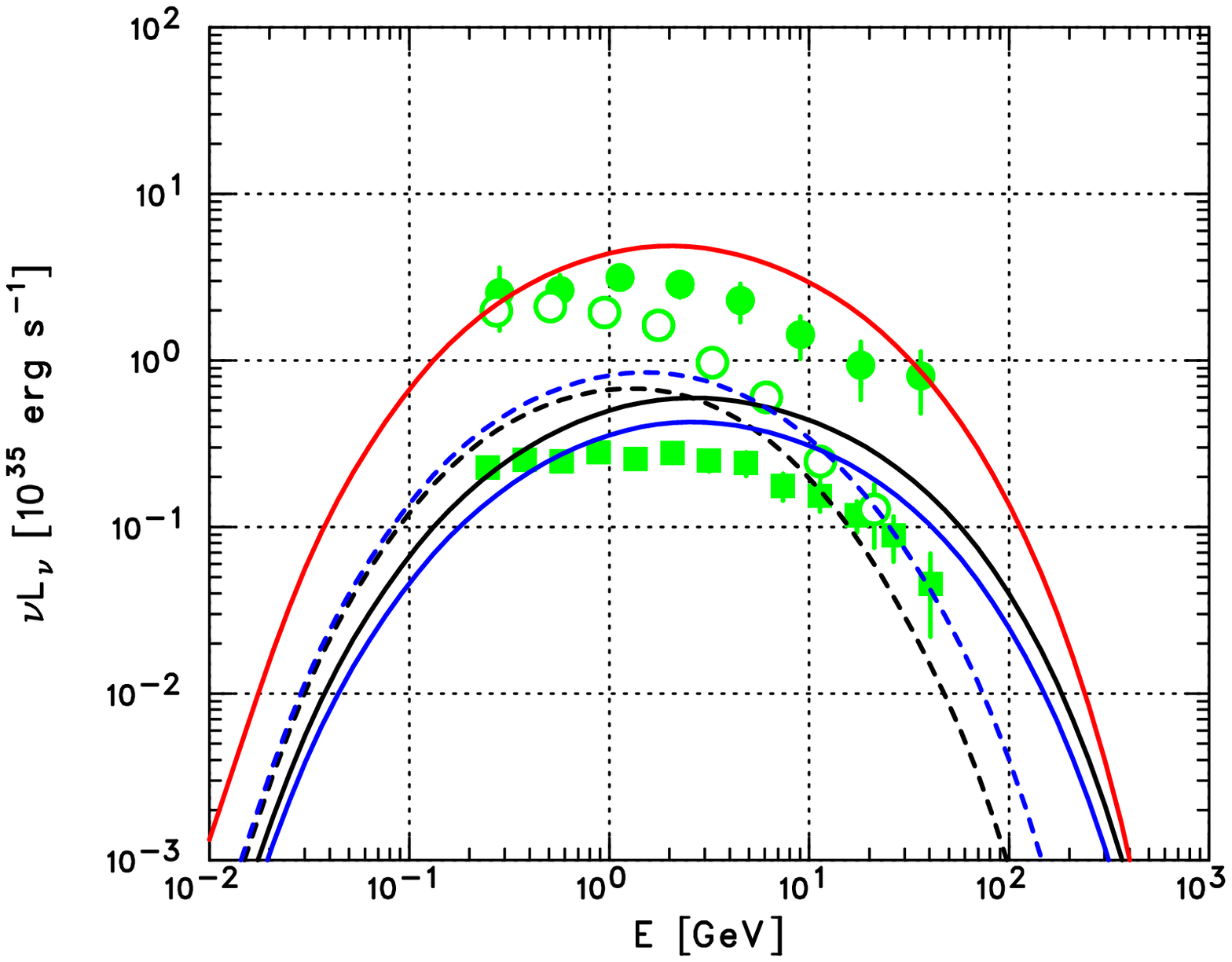}
\caption{\small 
$\pi^0$-decay $\gamma$-ray  spectra calculated for 
 the reacceleration model using various sets of parameters. 
Spectral data points are for SNR W51C \citep[filled circles:][]{FermiW51C}, 
W44 \citep[open circles:][]{FermiW44}, and IC 443 \citep[squares:][]{FermiIC443}.
}
\label{fig:Lumi}
\end{figure}
\begin{figure}[htbp] 
 \epsscale{1.15}
\plotone{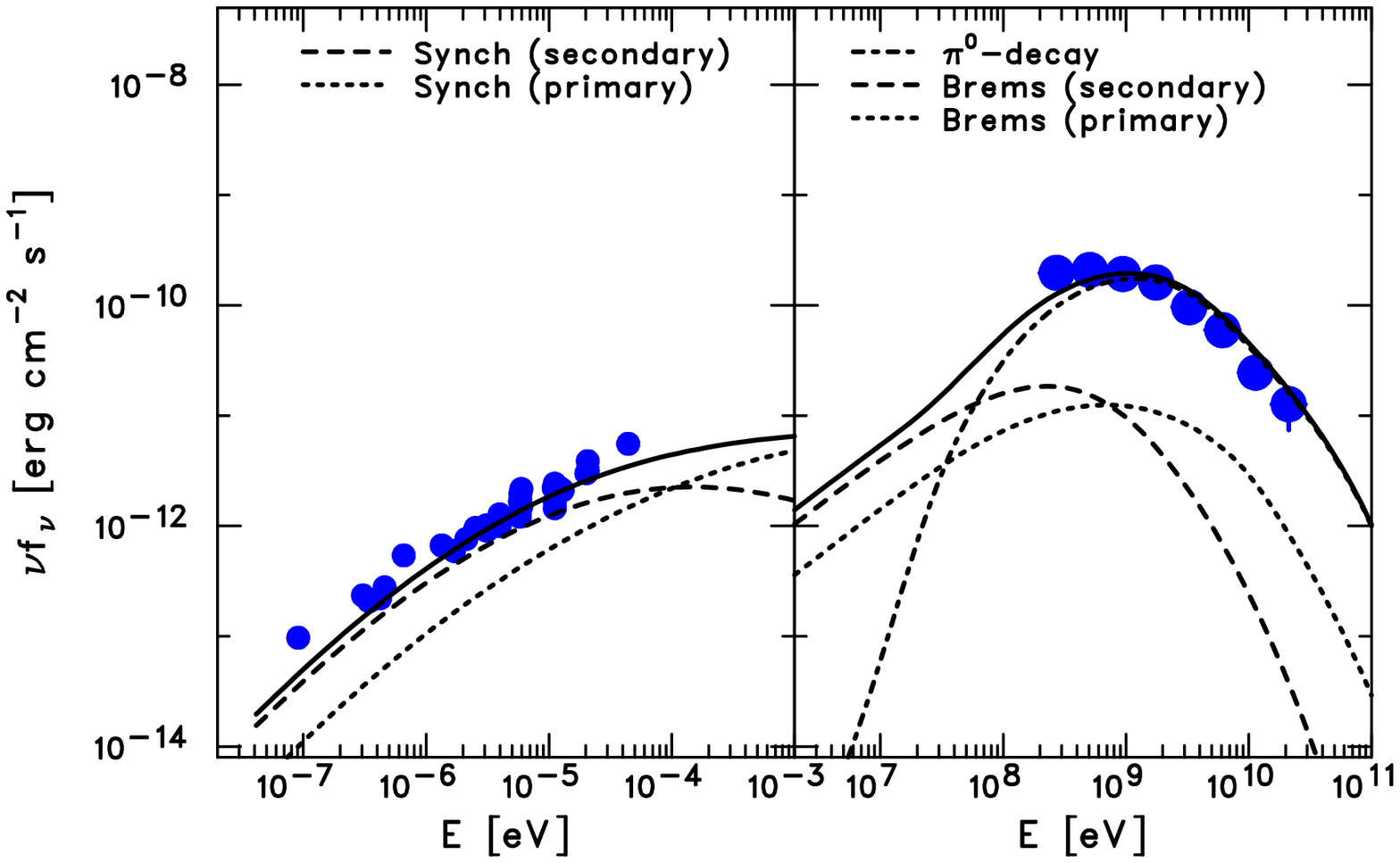}
\caption{\small 
Radio (left) and $\gamma$-ray (right) spectra of SNR W44 together with 
the reacceleration model using  
the parameters in Table \ref{tbl:model}, most of which are copied from 
 \citet{Reach05}. The radio fluxes are scaled by a factor of 0.5
(see the text).   }
\label{fig:W44_B}
\end{figure}

\section{Results}

Let us calculate the nonthermal radiation arising from the radiatively 
compressed clouds. 
The gas density and magnetic field strength are constant over the emission 
volume in which high-energy particles are distributed with the 
volume-integrated spectrum of $N(p,t_c)$. 
Leptonic components include synchrotron radiation, inverse-Compton scattering, 
and relativistic bremsstrahlung; the inverse-Compton emission is negligible 
as compared to the bremsstrahlung component   \citep[see, e.g.,][]{FermiW44}. 
Inelastic collisions between CR protons/nuclei  
and gas nuclei lead to $\pi^0$-decay $\gamma$-ray emission, 
which constitutes the main component in the \emph{Fermi}-LAT band 
under the assumption that only pre-existing CRs can be accelerated at the cloud shock. 

We do not consider the densest part of the interacting molecular cloud. 
For example, 
detections of OH(1720 MHz) maser emission \citep{OHmaser} 
indicate the presence of slow $C$-type shocks
in dense molecular clumps, say $v_{s7} \sim 0.3$ and $n_{0,2} \sim 10$, 
in addition to the fast $J$-type shock described in \S\ref{sec:model}. 
However, 
the $\gamma$-ray emission from the slow non-dissociative shocks 
is not expected to be strong 
because of inefficient shock-acceleration and weaker gas compression 
\citep[see also][]{Bykov00}.
Indeed,
the contribution from such dense molecular clumps to the total radio intensity 
appears small in SNR W44 \citep{Reach05}.
The largest preshock density we should consider is given by Eq.~(\ref{eq:n0}). 
Also, we ignore the blastwave region, since the radiative compression is 
essential in our model.

\subsection{Gamma-ray Luminosity}

We show here that the $\gamma$-ray luminosity anticipated 
within this scenario agrees  well with the observed luminosity of 
$\sim 10^{35}\ \rm erg\ s^{-1}$. 
In Figure~\ref{fig:Lumi}, the $\gamma$-ray spectra of SNRs W51C, W44, and IC~443 
measured with 
the \emph{Fermi}-LAT are shown in the so-called $\nu L_\nu$ form in units of 
$10^{35}\ \rm erg\ s^{-1}$. The LAT spectral points are taken from 
\citet{FermiW51C,FermiW44,FermiIC443} and converted into the $\nu L_\nu$ form 
using the distances of 6 kpc (W51C), 2.9 kpc (W44),  and 1.5 kpc (IC~443).

To demonstrate the expected level  of the $\gamma$-ray luminosity, 
we present the spectra of 
$\pi^0$-decay $\gamma$-rays with varying $R$, $n_0$, and $E_{51}$. 
The following parameters are fixed: $b=2$, $f=0.2$, $n_{a,0}=1$. 
Also, $p_{\rm max}$ is set by adopting $\eta =10$, 
and $p_{\rm br}$ is set by $T_4 =2$ and by the ionization fraction calculated 
based on HM89, here and hereafter. 
The $\gamma$-ray spectra simply scale as $\propto f$, and  depend very weakly 
on $n_{a}$. 
The black lines in Fig.~\ref{fig:Lumi} show the results 
obtained for $n_{0,2} = 0.3$ (solid curve) and $n_{0,2} = 3$ (dashed curve) 
in the case of $R=10\ \rm pc$ and $E_{51} =1$. 
The  $\gamma$-ray luminosity around 1 GeV varies only within a factor of 
$\sim 2$ between $n_{0,2} = 0.3$ and $n_{0,2} = 3$, while that at 100 GeV 
changes more than an order of magnitude. 
On the other hand, the blue lines in Fig.~\ref{fig:Lumi} show the spectra  
calculated for $R=5\ \rm pc$ (solid curve) and $R=15\ \rm pc$ (dashed curve) 
in the case of $n_{0,2} = 1$ and $E_{51} =1$. 
The $\gamma$-ray luminosity differs by a factor of $\sim 3$. 
We note that the shocked cloud mass amounts to $\sim 10^4\, M_\sun$ 
in the case of $R=15\ \rm pc$. 
Finally, 
to explore the most luminous scenario, we adopt $E_{51} =5$ together with 
$R=30\ \rm pc$ and $n_{0,2} = 1$ (red curve). 
The $\gamma$-ray luminosity reaches $\sim 10^{36}\ \rm erg\ s^{-1}$, 
in good agreement with the observations of SNR W51C, 
which is indeed the most luminous SNRs in gamma-rays. 
Our model generally predicts 
$L_{\gamma} \sim 10^{35}\, (f/0.2)\, E_{51}^{2/3}\ \rm erg\ s^{-1}$ 
to the first order, which led us to conclude that 
 the \emph{Fermi}-detected $\gamma$-rays 
are quite likely due to the decays of $\pi^0$-mesons produced by the 
 pre-existing CRs accelerated and subsequently compressed in the shocked cloud.

\subsection{Flat Radio Spectra} 

The GeV-bright SNRs W51C, W44, and IC~443 are also radio-bright objects. 
As shell-type SNRs, 
their radio spectra are remarkably flat  with spectral index
of  $\alpha \simeq 0.26$ \citep[W51C:][]{MK94}, 
$\alpha \simeq 0.37$ \citep[W44:][]{W44radio}, and 
$\alpha  \simeq 0.36$ \citep[IC 443:][]{Erickson85}, 
with a typical  uncertainty of 0.02, 
being inconsistent with $\alpha = 0.5$ that is expected by shock-acceleration theory. 
Our model naturally explains the flat radio spectrum. 
Let us demonstrate 
by presenting the radio and $\gamma$-ray modeling of SNR W44 
how  the radio and $\gamma$-ray spectra can be 
simultaneously reproduced. 

Radio \citep{W44radio} and $\gamma$-ray \citep{FermiW44} spectra of SNR W44 
are shown in Figure~\ref{fig:W44_B}.
One half of the total synchrotron flux measured for W44 
is assumed to originate in a fast molecular shock, 
which is roughly consistent with Table 2 of \citet{Reach05}. 
The rest is attributed to the blastwave region. 
Table~2 of \citet{Reach05} was chosen as an initial set of model parameters: 
$R_{12.5}=1$, $t_4=1$, $E_{51}=5$, and $n_{0,2} = 2$. 
We then attempted to reproduce the nonthermal radiation spectra by varying 
$f$ and $B_0$, and found that $f=0.18$ and  $B_0 = 25\, \mu$G provide a 
good fit to the data (see Table \ref{tbl:model} and Fig.~\ref{fig:W44_B}).
The radio measurements can be reconciled with this model, in which 
the synchrotron radiation is largely contributed by secondary electrons and positrons. 
The flat radio spectrum is generically expected in our model. 

\section{Discussion}

The radiatively-compressed cloud  provides a simple explanation for 
the radio and $\gamma$-ray data. 
Interestingly, the observed steepening in the $\gamma$-ray spectra 
is successfully reproduced by  $p_{\rm br}$ in the case of SNR W44. 
 However, there may be other explanations for the steepening. 
 For example, high-energy particles may be prone to escape from the 
 compressed magnetized cloud. 
 Also, the spectral break may be due to the fact that the crushed clouds 
 have a range of $n_0$. 
A superposition of $\gamma$-ray spectra characterized by different $p_{\rm max}$ 
(as a result of different $n_0$) could look like a break. 
  We are primarily interested in understanding the $\gamma$-ray luminosity 
 rather than the spectral shape in this paper, and therefore we did not explore  this issue. 

The simple re-acceleration of pre-existing CRs and subsequent compression alone would 
 not fully explain the $\gamma$-rays associated with cloud-interacting SNRs. 
For example, the GeV-TeV $\gamma$-ray emission found outside the radio 
boundary of SNR W28 \citep{HESSW28,AGILE_W28,FermiW28} may represent 
the molecular cloud illuminated by runaway CRs  \citep{AA96,Gabici09}. 
Also, we assumed pre-existing CRs in the cloud to have the same spectra as 
the galactic CRs in the vicinity of the solar system. 
However, the ambient CRs in the pre-shock cloud may deviate from the 
galactic pool due to the runaway CRs that have escaped from SNR shocks 
at earlier epochs. 
If this is the case, modeling of the $\gamma$-ray spectrum, at TeV energies in particular, 
should take into account modified pre-existing CRs in the pre-shock cloud. 

\acknowledgments

We acknowledge the useful suggestions of the anonymous referee, which 
improved the manuscript. 
We wish to thank Heinz V\"olk and Felix Aharonian for valuable discussions.


\end{document}